%% file: ms.tex
\documentclass[a4paper,conference]{IEEEtran}
\usepackage[left=1.57cm,right=1.57cm,top=0.95cm,bottom=2.7cm]{geometry}
\IEEEoverridecommandlockouts

\usepackage{url}
\usepackage[utf8]{inputenc}
\usepackage{amsmath}
\usepackage{bm}
\usepackage{amssymb}
\usepackage{xfrac}

\usepackage[acronyms,nonumberlist,nopostdot,nomain,nogroupskip]{glossaries}
\usepackage{tablefootnote}
\usepackage{cite}

\usepackage{tabularx}
\usepackage{booktabs}

\usepackage{comment}

\usepackage{tikz}
\usepackage{pgfplots}
\pgfplotsset{compat=newest}
\pgfplotsset{plot coordinates/math parser=false}
\newlength\fheight
\newlength\fwidth
\usetikzlibrary{plotmarks,patterns,decorations.pathreplacing,backgrounds,calc,arrows,arrows.meta,spy,matrix}
\usepgfplotslibrary{patchplots,groupplots}
\usepackage{tikzscale}
\usepackage{hyperref}

\newif\ifexttikz
\exttikzfalse

\ifexttikz
	\usetikzlibrary{external}
\fi

\usepackage{multirow}
\usepackage{enumerate}
\usetikzlibrary {patterns.meta}

\usepackage{mathtools}

\input{acronymsSorted.tex}
\input{myTikz.tex}

\definecolor{desireRed}{RGB}{230,57,60}%
\definecolor{darkPurple}{RGB}{59,31,43}%
\definecolor{springGreen}{RGB}{37,223,145}%
\definecolor{queenBlue}{RGB}{69,123,157}%
\definecolor{spaceCadet}{RGB}{29,53,87}%

\definecolor{primaryColor}{HTML}{F06449}
\definecolor{secondaryColor}{HTML}{5BC3EB}
\definecolor{tertiaryColor}{HTML}{36382E}

\usepackage[font=footnotesize]{caption}
\usepackage[font=footnotesize]{subcaption}

\def\BibTeX{{\rm B\kern-.05em{\sc i\kern-.025em b}\kern-.08em
    T\kern-.1667em\lower.7ex\hbox{E}\kern-.125emX}}
    
\makeglossaries    
    
\begin{document}

\title{End-to-End Simulation of \\ 5G Networks Assisted by IRS and AF Relays}

\author{\IEEEauthorblockN{
Matteo~Pagin\IEEEauthorrefmark{1},
Marco~Giordani\IEEEauthorrefmark{1},
Amir~Ashtari~Gargari\IEEEauthorrefmark{1},
Alberto~Rech\IEEEauthorrefmark{1},
Federico~Moretto\IEEEauthorrefmark{1},\\
Stefano~Tomasin\IEEEauthorrefmark{1}, 
Jonathan~Gambini\IEEEauthorrefmark{3}, 
Michele~Zorzi\IEEEauthorrefmark{1}}\vspace{2mm}

  \IEEEauthorblockA{\IEEEauthorrefmark{1}Department of Information Engineering, University of Padova, Italy. Email:\texttt{\{name.surname\}@dei.unipd.it}\\
  \IEEEauthorrefmark{3}Milan Research Center, HUAWEI, Italy. Email:\texttt{jonathan.gambini@huawei.com}
  }
}

\maketitle

\begin{abstract}
The high propagation and penetration loss experienced at \gls{mmwave} frequencies requires ultra-dense deployments of \gls{5g} base stations, which may be infeasible and costly for network operators. 
\Gls{iab} has been proposed to partially address this issue, even though raising concerns in terms of power consumption and scalability.
Recently, the research community has been investigating \glspl{irs} and \gls{af} relays as more energy-efficient alternatives to solve coverage issues in 5G scenarios.
Along these lines, this paper relies on a new simulation framework, based on ns-3, to simulate IRS/AF systems with a full-stack, end-to-end perspective, with considerations on to the impact of the channel model and the protocol stack of 5G NR networks.
Our goal is to demonstrate whether these technologies can be used to relay 5G traffic requests and, if so, how to dimension IRS/AF nodes as a function of the number of end users.
\end{abstract}

\begin{picture}(0,0)(0,-350)
\put(0,0){
\put(0,0){\qquad \qquad \quad This paper has been submitted to IEEE for publication. Copyright may change without notice.}}
\end{picture}

\glsresetall


\section{Introduction}
\label{sec:intro}


\Gls{5g} networks are being rolled out worldwide 
as a means to provide $20\times$ higher peak throughput and $10\times$ lower latency than previous generations.
To accomplish this, the \gls{3gpp} has released a new set of innovations for 5G networks~\cite{38300}, including the support for network operations in the \gls{mmwave} spectrum, in combination with \gls{m-mimo} technologies.
Transmissions at \glspl{mmwave}, in turn, introduce several propagation issues, first and foremost the severe path and penetration losses, which force the communication to be in short range~\cite{rangan2017potentials}.
A possible solution could lie in a denser deployment of 5G \gls{mmwave} base stations, which however would be costly for network operators, especially in terms of sites acquisition campaigns, rental fees, and fiber optic layout to provide wired~backhauling~\cite{lopez2015towards}.

To solve this issue, the 3GPP approved, as part of its 5G NR specifications for Rel-16~\cite{38874}, \gls{iab} as a new paradigm to replace fiber-like infrastructures with self-configuring relays operating through wireless (\gls{mmwave}) backhaul links.
Despite this potential, \mbox{\gls{m-mimo}-assisted} IAB still requires complex signal processing as well as costly and energy consuming hardware~\cite{polese2020integrated}.
This issue is exacerbated in rural/remote areas, where harsh weather and terrain, and the lack of a powerful electrical grid in many cases, may further complicate IAB installation~\cite{chaoub2021bridging}.

In light of this, new technologies based on \glspl{irs} and \gls{af} relays have been proposed as promising alternatives to overcome the coverage issues of \gls{mmwave} networks, with energy efficiency in mind~\cite{flamini2022towards}. 
An \gls{irs} is a meta-surface that can be programmed to favorably alter an \gls{em} field towards an intended destination. 
Specifically, \glspl{irs} are nodes which passively beamform the impinging signal, without amplification, thus being able to guarantee minimum capacity requirements in dead spots with lower power consumption compared to IAB~\cite{bjornson2019intelligent}. 
\Gls{af} relays, instead, are envisioned to capture an incident electromagnetic wave 
coming from a base station, to actively amplify the received signal, and to re-radiate it 
towards a target area to be served. They are candidates for achieving higher capacity with respect to IRS nodes, at the expense of higher cost and amplification noise~\cite{huang2019reconfigurable}.

Whether these technologies will be able to fulfill 5G (and beyond) service requirements and, if so, how to properly dimension IRS/AF systems, are still crucial issues that remain unsolved. 
While field experiments with real hardware are infeasible due to scalability and flexibility concerns, as well as the high cost of testbed components, computer-based simulations represent a viable approach for testing and calibrating IRS/AF deployments. 
Prior works, e.g.,~\cite{wu2018intelligent,9282349}, have addressed this task, though focusing on link-level analyses, which typically adopt conservative assumptions on the system architecture, and should be taken as a lower bound for more representative end-to-end performance studies.

To fill this gap, in this paper we provide a more comprehensive system-level performance evaluation of IRS/AF deployments using a new simulation
framework that operates end-to-end, thus incorporating the interplay with the \gls{5g} NR protocol stack and relative control tasks, as well as the impact of the upper (including transport and application) layers.
Our framework is based on \mbox{ns-3}~\cite{ns3}, an open-source discrete-event simulator for wireless networks.
Specifically, we describe our ns-3 implementation of the IRS/AF channel, based on the current 3GPP channel model for 5G networks standardized in~\cite{3gpp.38.901} and implemented, e.g., in the \texttt{ns3-mmwave} module~\cite{mezzavilla2018end}, which models the \gls{phy} and \gls{mac} layers of the \gls{5g} NR protocol stack.
Based on this, we conduct an extensive simulation campaign to study the performance of IRS/AF nodes for relaying connectivity requests from end users, compared to a baseline solution in which relays are not deployed. 
We demonstrate that IRSs and AF relays are valid solutions, especially in small networks, even though high-EIRP AF relays are required to support more aggressive traffic applications.
Based on our simulations, we provide guidelines towards the optimal dimensioning of IRS and AF configurations, in terms of number of antenna elements and amplification power.

The rest of the paper is organized as follows. 
In Sec.~\ref{sec:ch_model_ext} we present a mathematical characterization of the channel for IRS/AF relays, based on the 3GPP channel model for 5G networks.
Sec.~\ref{sec:simulator} describes our simulation methodology for IRS/AF relays. 
In Sec.~\ref{sec:results} we show our main numerical results, while Sec.~\ref{sec:conclusion} concludes the work with suggestions for future research.

\section{A 3GPP TR 38.901-Based Signal Model for IRS/AF-Assisted 5G Networks} 
\label{sec:ch_model_ext}

A realistic characterization of the channel is the first step to obtain accurate simulation results. 
Therefore in this section we provide a mathematical model for the IRS and \gls{af} relay channels  (Secs.~\ref{sec:irs_phy_model} and~\ref{sec:af_phy_model}, respectively), based on the standard 3GPP channel model for 5G networks (Sec.~\ref{sec:baseline_model}). 

\emph{Notation.} We use boldface upper- and lower-case letters to refer to matrices and vectors, respectively, while lower-case letters denote scalars. We use $\bm{I}_{N}$ to denote the identity matrix of order $N$, $[ \bm{\Phi} ]_{j, k}$ to indicate the $(j, k)$-th
entry of matrix $\bm{\Phi}$, $\mathrm{diag} ( \phi_1,\dots, \phi_N) $ to indicate an  $N$~$\times$~$N$ diagonal matrix with entries $\{\phi_j \, | \, j=1, \dots, N \}$. We use the superscripts T, H and~$*$ for transposition, Hermitian transposition, and conjugation, respectively.



\subsection{The TR~38.901 Channel Model for 5G NR}
 \label{sec:baseline_model}

We consider the 3GPP TR~38.901 \gls{scm}, standardized in~\cite{3gpp.38.901}. 
This choice is motivated by the fact that TR~38.901 supports a wide range of frequencies, from $0.5$ to $100$ GHz, and can be integrated with realistic beamforming models. Furthermore, it is suggested and adopted by the \gls{3gpp} itself for the performance evaluation of \gls{5g} networks via system-level simulations.

In particular, the TR~38.901 model outlines the procedures for generating a channel matrix $\bm{H}$ whose entries $\bm{H}_{p, q} (t, \tau)$ correspond to the impulse response of the channel between the $p$-th radiating element of the antenna array of the signal source (S), and the $q$-th radiating element of the antenna array of its destination (D), at time $t$ and with delay $\tau$. 
To model multipath fading, each of these terms is computed as the superposition of $N$ different clusters, each of which consists of $M$ rays that arrive (depart) to (from) the antenna arrays with specific angles and powers. Based on~\cite{3gpp.38.901}, and using the simplifications proposed in~\cite{zugno2020implementation}, the generic entry $\bm{H}_{p, q} (t, \tau)$ of the channel matrix can then be computed as:

\begin{equation}
\label{eq:ch_model_full}
\begin{aligned}
\bm{H}_{p, q}(t, \tau)=& \sum_{n=1}^{N} \sqrt{\frac{P_{n}}{M}} \sum_{m=1}^{M} \overline{\mathbf{F}}_{r x}\left(\theta_{n, m}^{A}, \phi_{n, m}^{A}\right) \\
& \times\left[\begin{array}{cr}
e^{j \Phi_{n, m}^{\theta, \theta}} & \sqrt{K_{n, m}^{-1}} e^{j \Phi_{n, m}^{\theta, \phi}} \\
\sqrt{K_{n, m}^{-1}} e^{j \Phi_{n, m}^{\phi, \theta}} & e^{j \Phi_{n, m}^{\phi, \phi}}
\end{array}\right] \\
& \times \overline{\mathbf{F}}_{tx}\left(\theta_{n, m}^{D}, \phi_{n, m}^{D}\right) \\
& \times e^{j \overline{\mathbf{k}}_{rx, n, m}^{T} \overline{\mathbf{d}}_{rx, p} e^{j \overline{\mathbf{k}}_{tx, n, m}^{T} \overline{\mathbf{d}}_{tx, q}}} \\
& \times e^{j 2 \pi v_{n} t} \delta\left(\tau-\tau_{n}\right).
\end{aligned}
\end{equation}
For a complete description of the specific terms appearing in Eq.~\eqref{eq:ch_model_full} we refer the interested reader to~\cite{zugno2020implementation}.

Then, a frequency-flat path gain term is added to each channel coefficient as a function of the carrier frequency $f_c$ and the distance $d$ between the endpoints, i.e.,  
\begin{equation}
\text{PL} (d, f_c) = A \log_{10} (d) + B + C \log_{10} (f_c) + X \, [\mathrm{dB}],
\label{eq:pl}
\end{equation}
where model parameters $A, B$ and $C$ depend on the propagation conditions and the type of environment, and $X$ is an optional term for representing shadowing~\cite{zugno2020implementation}. 

\begin{figure}[t]
  \centering
    \includegraphics[width=0.48\textwidth]{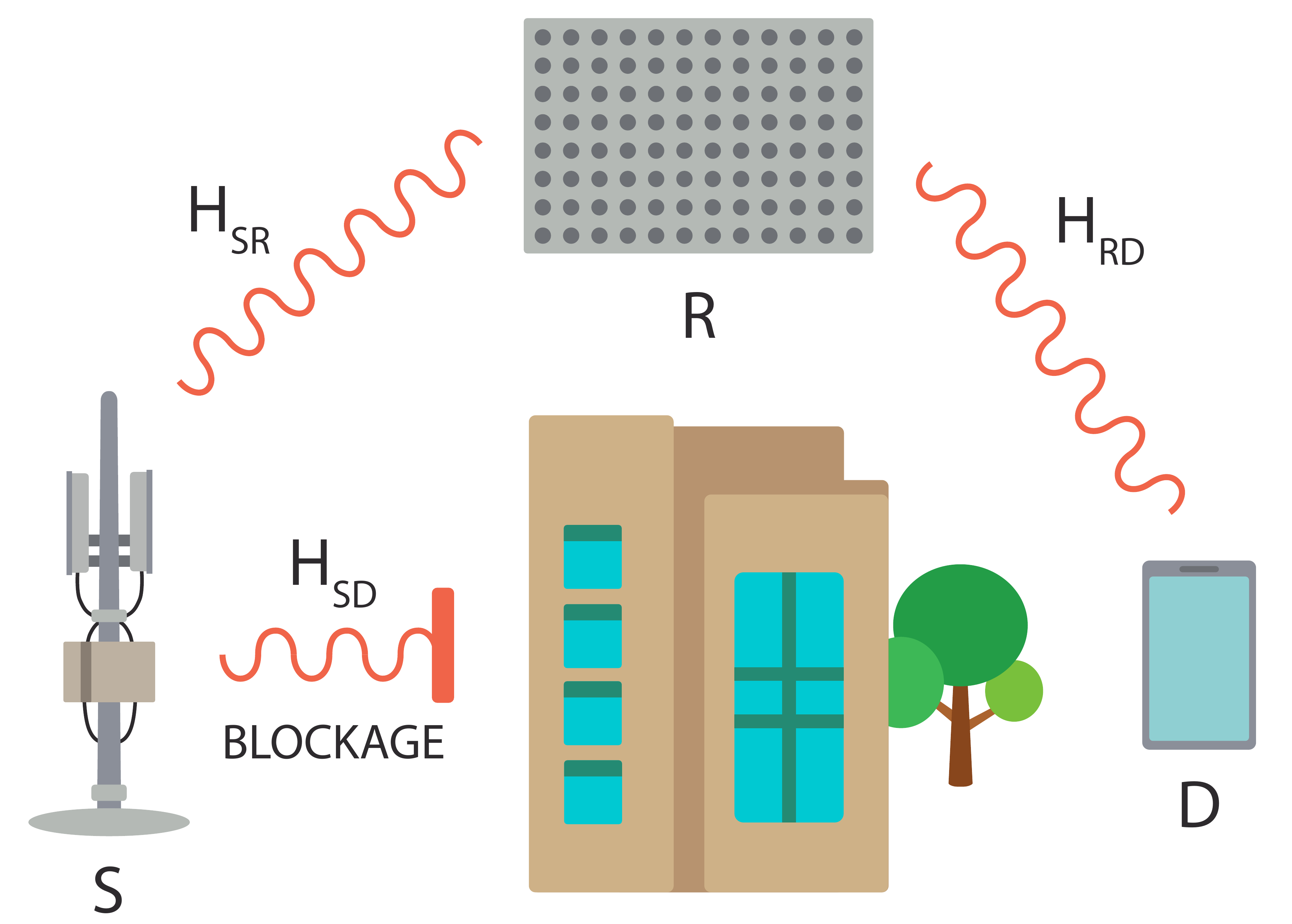}
  \caption{A typical urban scenario where a relay (R) can be used to bridge the signal from a source (S) to a destination (D), that would otherwise communicate in \acrshort{nlos}, i.e., the direct link between S and D is blocked due to obstacles such as buildings and/or vegetation.}
  \label{Fig:scenario_high_lvl}
\end{figure}

We consider the transmission of a single data stream $x_{\rm S}$, i.e., a sequence of signals, from a source S to a destination D via a relay R, as depicted in Fig.~\ref{Fig:scenario_high_lvl}. 
Then, the channel matrix is combined with the beamforming vectors used at S and D, in order to obtain the \gls{sinr} experienced at D. 
In particular, let $x_{\rm S}$ be the signal transmitted from S to D, and $\bm{w}_{\rm S}$, $\bm{w}_{\rm D}$ and $\bm{w}_{\rm I}$ be the beamforming vectors used at S, D and the I-th interferer, respectively. Moreover, we define the following matrices:
$\bm{H}_{\rm SD}$ is the channel matrix between the source and the destination,
 $\bm{H}_{\rm ID}$ is the channel matrix between the \mbox{I-th} interferer and the destination,
 $\bm{H}_{\rm IR}$ is the channel matrix from the \mbox{I-th} interferer to the relay,
 $\bm{H}_{\rm SR}$ is the channel matrix between the source and the relay,
 and $\bm{H}_{\rm RD}$ is the channel matrix between the relay and the destination.
In a relay-free environment, the signal received at the \gls{ue} is computed as:
\begin{equation}
    \label{eq:input_output_nodev}
    y_{\rm D} = \bm{w}_{\rm D}^{T} \bm{H}_{\rm SD} \bm{w}_{\rm S} x_{S} + \sum_{\rm{I}=1}^{N} \bm{w}_{\rm D}^{T} \bm{H}_{\rm ID}  \bm{w}_{\rm I} x_{\rm I} + \bm{w}_{\rm D}^{T} \bm{n}_{\rm D}.
\end{equation}
where $\bm{n}_D$ represents the circularly symmetric complex Gaussian noise vector with correlation matrix $\sigma_{\rm N}^2 \bm{I} $, and $\bm{w}_{\rm D}^{T} \bm{H}_{\rm ID} \bm{w}_{\rm I} x_{\rm I}$ is the signal received from the $\rm I$-th interferer.
Accordingly, the \gls{sinr} at D reads:
\begin{equation}
	\Lambda = \frac{ \lVert \bm{w}_{ \textrm D}^{T} \bm{H}_{\textrm {SD}}\bm{w}_{ \textrm S} \rVert^2 \sigma_{\rm S}^2 } { \sum_{\rm{I}=1}^{N} \lVert \bm{w}_{\rm D}^{T} \bm{H}_{\rm ID} \bm{w}_{\rm I} \rVert^2 \sigma_{\rm I}^2 + \sigma_{\rm N}^2 },
\end{equation}
where $\sigma_{\rm S}^2$ and $\sigma_{\rm I}^2$ are the powers of the useful and the $\rm{I}$-th interfering signals, respectively.
\subsection{A Signal Model for the IRS}
\label{sec:irs_phy_model} 
An \gls{irs} is a planar surface made of $N_{\rm R}$ low-cost passive reflecting elements that can be programmed to alter an \gls{em} field, for example to achieve three-dimensional beamforming towards an intended destination.
The working principle is similar to that of a conventional relay, the main difference being that while the latter amplifies the received signal before retransmitting it, an \gls{irs} reflects and beamforms the signal without introducing any amplification, thus saving power compared with other relaying solutions~\cite{bjornson2019intelligent}. 


In particular, each element of the \gls{irs} acts as an 
antenna that captures and reflects the incoming signals, introducing a phase shift on the baseband-equivalent signal. We denote with $\phi_n=e^{j\theta_n}, \, n = 1, \ldots, N_{\rm R}$, the reflection coefficient of the $n$-th \gls{irs} element, where $\theta_n \in [-\pi,\pi ] $ is the induced, controllable phase shift. 
Adopting a complex baseband notation, the signal $\bm{z} \in {\mathbb C}^{N_{\rm R} \times 1} $ reflected by an IRS (denoted as R), impinged with a signal $x_{\rm S}$ originating from a source S, reads
\begin{equation}
\bm{z}=\bm{\Phi}\bm{H}_{\rm SR} \bm{w}_{\rm S} x_{\rm S} ,
\end{equation}
where $\bm{\Phi}$ is a diagonal matrix defined as $\bm{\Phi} \doteq \mathrm{diag} (\phi_1,\ldots,\phi_{N_{\rm R}})$, and typically referred to as {\em \gls{irs} configuration}. 
Therefore, the signal received at the intended destination D (under far-field assumption with respect to the \gls{irs}) can be expressed as
\begin{equation}
	\label{eq:irs_iput_output}
    y_{\rm D} = \bm{w}_{\rm D}^{T} \bm{H}_{\rm RD} \bm{\Phi} \bm{H}_{\rm {SR}}  \bm{w}_{\rm S} x_S + \bm{w}_{\rm D}^{T} \bm{H}_{\rm SD} \bm{w}_{\rm S} x_S + \bm{w}_{\rm D}^{T} \bm{n}_{\rm D}.
\end{equation}
 
 \subsection{A Signal Model for the AF Relay}
 \label{sec:af_phy_model}
 
  \Gls{af} relays have been studied in the context of cooperative communications as a means to regenerate a relayed signal through amplification, with the goal of improving the system capacity. 
  Unlike \glspl{irs}, \gls{af} relays feature a non-negligible power consumption, and introduce noise amplification.

In this work we consider as \gls{af} relay a device equipped with $M_{\rm R}$ transmit and $M_{\rm R}$ receive antennas.
Therefore, the signal received at D is:
\begin{align}
\label{eq:af_iput_output}
    y_{\rm D} = \,\, &\bm{w}_{\rm D}^{T} \bm{H}_{\rm RD} \bm{\Phi} \bm{H}_{\rm {SR}} \bm{w}_{\rm S} x_{\rm S} + \bm{w}_{\rm D}^{T} \bm{H}_{\rm SD} \bm{w}_{\rm S} x_{\rm S}  \nonumber \\
   &+ \bm{w}_{\rm D}^{T} \bm{H}_{\rm RD} \bm{\Phi} \bm{n}_{\rm R} + \bm{w}_{\rm D}^{T} \bm{n}_{\rm D},
\end{align}
where in this case matrix $\bm{\Phi}$ also accounts for the amplification gain, and its structure depends on the specific relay design.
Moreover, $\bm{n}_{\rm R}$ represents the circularly symmetric complex Gaussian noise vector with covariance matrix $\sigma_{\rm N_R}^2 \bm{I}_{M_{\rm R}} $.
Then, the power of the noise term relayed by the \gls{af} relay to receiver D and measured after the combiner at the \gls{ue}, is
\begin{equation} 
\label{eq:af_noise_power}
    \begin{split}
        \hat{\sigma}_{\rm N_R}^2 &= \left( \bm{w}_{\rm D}^{T} \bm{H}_{\rm RD} \bm{\Phi} \right) \left( \bm{w}_{\rm D}^{T} \bm{H}_{\rm RD} \bm{\Phi} \right)^{H} \sigma_{\rm N_R}^2 \\
         & = \bm{w}_{\rm D}^{T} \bm{H}_{\rm RD} \bm{\Phi} \bm{\Phi}^{H} \bm{H}_{\rm RD}^{H} {\bm w}^{*}_{\rm D} \sigma_{\rm N_R}^2.
    \end{split}
\end{equation}


\section{A Full-Stack Simulator for IRS/AF Relays}
\label{sec:simulator}

Despite the availability of accurate sub-6 GHz and \gls{mmwave} channel models, analytical evaluations of the \gls{5g} NR protocol stack introduce several assumptions in the system architecture, and are generally not desirable~\cite{gkonis2020comprehensive}. 
Additionally, 5G/6G cellular networks are rapidly shifting towards open and controllable network configurations, which further introduce unprecedented data-driven programmability~\cite{bonati2020open}. 
In these regards, computer simulators are emerging as a valuable tool to let researchers better understand the performance of wireless networks, and dimension them accordingly~\cite{wilhelmi2021usage}.  

Several simulators for 5G cellular and vehicular networks are available in the literature~\cite{mezzavilla2018end,choi20195g, nardini2020simu5g, patriciello2019e2e, pratschner2018versatile, muller2018flexible, jao2018wise,drago2020millicar}. However, they provide a detailed characterization of either the lower (i.e., at link-level) or the upper (i.e., at system-level) layers of the \gls{5g} NR protocol stack. 
Notably, the latter sacrifice \gls{phy} layer accuracy to reduce the computational complexity, but incorporate accurate models of the remainder of the protocol stack, thus enabling scalable end-to-end simulations.
Despite the many software-based evaluation platforms available, to the best of our knowledge there are no end-to-end simulators for IRSs and AF relays. In~\cite{polese2018end}, the authors presented an open-source module for \gls{iab}, even though it was not extended to support passive relays like IRSs. 
Moreover, the authors in~\cite{heimann2021modeling} presented an ns-3 \gls{irs} module, but their work focused on vehicular networks, and did not consider the case of AF~relays. 

In this paper we close the gap and propose an ns-3-based simulator for IRSs and AF relays.
Arguably, the main effect of the presence of these entities is the alteration of the wireless channel between the communication endpoints. 
Accordingly, our simulator extends the \texttt{ns-3 mmwave} module~\cite{mezzavilla2018end} (among the most popular 5G-oriented NR-compliant frameworks to simulate 5G networks) by implementing a new signal model for IRS and AF relays, following the characterization in Secs.~\ref{sec:irs_phy_model} and \ref{sec:af_phy_model}, respectively, which is then used to compute the \gls{sinr} experienced by signals transmitted over a relayed wireless link. 


\subsection{Implementation of the IRS/AF Signal Model} 
\label{sec:ext_ch_model}
 
In line with~\cite{zugno2020implementation}, we assume that the transmission of the signal $x_{\rm S}$ occurs over a frequency-selective wireless channel as \gls{5g} NR supports network operations with a bandwidth up to 400 MHz, when using FR2~\cite{38101_1}. 
Therefore, the evaluation of the \gls{sinr} requires, among other things, the computation of the \gls{psd} of the useful component of the signal at D, i.e., $\mathcal{P}_{r x}$, starting from that of the input signal $\mathcal{P}_{t x}$. Additionally, we consider that both the transmitter and the receiver feature \gls{m-mimo} arrays equipped with multiple antenna elements, and use the beamforming vectors $\bm{w}_{\textrm S}$ and $\bm{w}_{\textrm D}$, respectively.
Under these assumptions, the input-output relationship in~\eqref{eq:input_output_nodev} becomes~\cite{bjornson2019intelligent, wu2019intelligent}:
\begin{equation}
\label{eq:mimo_input_output_relay}
\begin{aligned}
	y_{D} = \,\, & \bm{w}_{\rm D}^{\rm T} \bm{H}_{\rm RD} \bm{\Phi} \bm{H}_{\rm {SR}} \bm{w}_{\rm S} x_{S} + \bm{w}_{\rm D}^{T} \bm{H}_{\rm SD} \bm{w}_{\rm S} x_{S} + \tilde{n} + \\
	& \sum_{\rm{I}=1}^{N} \bm{w}_{\rm D}^{T} \bm{H}_{\rm {RD}} \bm{\Phi} \bm{H}_{\rm {IR}}  \bm{w}_{\rm I} x_{\rm I} + \sum_{\rm{I}=1}^{N} \bm{w}_{\rm D}^{T} \bm{H}_{\rm ID}  \bm{w}_{\rm I} x_{\rm I},
\end{aligned}
\end{equation}
where in turn $\tilde{n}$ is defined as:
\[ 
\tilde{n} = 
\begin{cases}
    \bm{w}_{\rm D}^{T} \bm{n}_{\rm D}		& \text{if IRS}, \\
    \bm{w}_{\rm D}^{T} \bm{n}_{\rm D} + \bm{w}_{\rm D}^{T} \bm{H}_{\rm RD} \bm{\Phi} \bm{n}_{\rm R}      & \text{if AF},
\end{cases}
\]
where matrix $\bm{\Phi}$ is the relay matrix, i.e., a matrix which fully encodes the effect of the relay, i.e., either IRS or AF, as described in Secs.~\ref{sec:irs_phy_model} and~\ref{sec:af_phy_model} for the single user case, respectively, over the wireless channel. 
Notably, S and D are either in \gls{nlos} (in this case they communicate via the relay, and we consider the direct link towards D to be unavailable), or in \gls{los} (in this case they do not use the relay). 
Accordingly, assuming that the source of interest is in NLOS with respect to its intended destination,~\eqref{eq:mimo_input_output_relay} becomes:
\begin{equation}
\label{eq:mimo_input_output_relay_reduced}
\begin{aligned}
	y_{D} = \,\, & \bm{w}_{\textrm D}^{\textrm T} \bm{H}_{\textrm {RD}} \bm{\Phi} \bm{H}_{\textrm {SR}} \bm{w}_{\textrm S} x_{S} + \sum_{\hat{I} \in  I_{\textrm{LOS}}} \bm{w}_{\rm D}^{T} \bm{H}_{\rm \hat{I} D}  \bm{w}_{\rm \hat{I}} x_{\rm \hat{I}} \\
 & + \sum_{ \bar{I} \in I_{\textrm{NLOS}}} \bm{w}_{\rm D}^{T} \bm{H}_{\rm{RD}} \bm{\Phi} \bm{H}_{\rm{ \bar{I} R}}  \bm{w}_{\rm \bar{I}} x_{\rm \bar{I}} + \tilde{n},
\end{aligned}
\end{equation}
where $I_{\textrm{LOS}}$ and $I_{\textrm{NLOS}}$ are the two disjoint sets of interferers which experience either a \gls{los} or a \gls{nlos} channel towards D, respectively.
Then, the \gls{psd} of the useful component of the signal at the receiver can be written as:
\begin{equation}
\label{eq:psd_relay}
\mathcal{P}_{r x}(t, f) = \mathcal{P}_{t x}(t, f) \lVert \bm{w}_{\textrm D}^{\textrm T} \bm{H}_{\textrm {RD}} \bm{\Phi} \bm{H}_{\textrm {SR}} \bm{w}_{\textrm S} \rVert^2.
\end{equation}

Based on the above definitions, our simulator computes the \gls{psd} by checking whether the communication from S to D involves a relay. 
If so, the \gls{psd} is computed according to the following steps.

\begin{enumerate}
    \item \emph{Channel matrices generation.} 
After having identified S and D as the two endpoints of the communication, the channel matrices $\bm{H}_{\textrm {SR}}$ and $\bm{H}_{\textrm {RD}}$ are computed based on~\eqref{eq:ch_model_full}~\cite{3gpp.38.901}. 

\item \emph{Configuration of the relay and the beamforming vectors.}
We assume that the choice of the beamforming vectors for both S and D ($\bm{w}_{\textrm D}$ and $\bm{w}_{\textrm S}$), as well as the relay configuration ($\bm{\Phi}$), consist in the choice of a \emph{codeword} from a pre-defined pre-computed \emph{codebook}. Moreover, we assume that the devices do not have full channel knowledge, i.e., they do not know the realizations of $\bm{H}_{\textrm {SR}}$ and $\bm{H}_{\textrm {RD}}$. 
Then, in line with the \gls{5g} NR beam management procedure~\cite{giordani2018tutorial}, the choice of the codeword in the codebook is performed via exhaustive search, i.e., by repeatedly sending pilot signals, and measuring the \gls{sinr} experienced with various configurations of the codebook. Eventually, we choose the combination of $\bm{w}_{\textrm D}$, $\bm{w}_{\textrm S}$, and $\bm{\Phi}$ yielding the highest \gls{sinr}.

Notably, this procedure is not repeated at each transmission opportunity. Instead, $\bm{w}_{\textrm D}$, $\bm{w}_{\textrm S}$, and $\bm{\Phi}$ are stored and re-used for the whole channel coherence time, to mimic the actual 5G NR beam management procedure, and also reduce the complexity of the simulations. 
Furthermore, the evaluation of the \gls{sinr} is performed by neglecting the small-scale fading terms, to further reduce the overhead. The small-scale fading will be eventually incorporated in Step 4 of the model.

\item \emph{Long-term computation.} 
Along the lines of~\cite{zugno2020implementation}, the \gls{psd} of the transmitted signal $x_S$ at D can be expressed as:
\begin{equation}
\begin{aligned}
&\mathcal{P}_{r x}(t, f) = \\
& = \mathcal{P}_{t x}(t, f) \lVert \bm{w}_{\textrm D}^{\textrm T} \bm{H}_{\textrm {RD}} \bm{\Phi} \bm{H}_{\textrm {SR}} \bm{w}_{\textrm S} \rVert^2 \\
&=\mathcal{P}_{t x}(t, f) \lVert \bm{w}_{\textrm D}^{\textrm T} \bm{H}_{\textrm {SRD}} \bm{w}_{\textrm S} \rVert^2 \\
& = \mathcal{P}_{t x}(t, f) \left\lVert \sum_{d=1}^{N_D} \sum_{s=1}^{N_S}  w^{\textrm D}_{d} h^{\rm SRD}_{d, s} (t, f)  w^{\textrm S}_{s} \right\rVert^2.
\end{aligned}
\label{eq:p-rx-sdr}
\end{equation}

In Eq.~\eqref{eq:p-rx-sdr}, $\bm{H}_{\rm SRD}$ is the equivalent channel matrix between S and D, whose generic entry  $h^{\rm SRD}_{d, s} (t, f)$ is:
\begin{equation}
\begin{aligned}
\label{eq:psd_relay_sums}
h^{\rm SRD}_{d, s} (t, f) &= \left[ \bm{H}_{\textrm {RD}} (t, f) \bm{\Phi} \bm{H}_{\textrm {SR}} (t, f) \right]_{d, s} \\
& = \sum_{n=1}^{N_{\rm RD}} \sum_{m=1}^{N_{\rm SR}} \sum_{k=1}^{N_{R}} \sum_{l=1}^{N_{\rm R}} h^{\rm RD}_{d, k, n} \, \phi_{k, l} \, h^{\rm SR}_{l, s, m} \\
& \quad \times e^{j 2 \pi v_{n} t} e^{j 2 \pi \tau_{n} f} \\
& \quad \times e^{j 2 \pi v_{m} t} e^{j 2 \pi \tau_{m} f},
\end{aligned}
\end{equation}
where $N_{\rm RD}$ and $N_{\rm SR}$ are the number of multipath clusters in $\bm{H}_{\textrm {RD}}$ and $\bm{H}_{\textrm {SR}}$, respectively. Moreover, $w^{\textrm S}_{s}$ and $w^{\textrm D}_{d}$ denote entries $s$ and $d$ of vectors $\bm{w}_{\rm S}$ and $\bm{w}_{\rm D}$, respectively.
Then, Step 3 consists in the evaluation of the long-term fading:
\begin{equation}
L_{n, m} \doteq \sum_{d=1}^{N_{\rm D}} \sum_{s=1}^{N_{\rm S}} \sum_{k=1}^{N_{\rm R}} \sum_{l=1}^{N_{\rm R}} w^{\rm D}_{d} \, h^{\rm RD}_{d, k, n} \, \phi_{k, l} \, h^{\rm SR}_{l, s, m} \, w^{\rm S}_{s}. 
\end{equation}

\item \emph{Small-scale fading and path loss.}
The small-scale fading terms are combined with the terms $L_{n, m}$ to compute the overall fading component of the \gls{psd} of interest:
\begin{equation}
	\mathcal{\tilde{P}}_{rx}(t, f) = \mathcal{P}_{t x}(t, f) \left\lVert \sum_{n=1}^{N_{\rm RD}} \sum_{m=1}^{N_{\rm SR}} L_{n, m} E_{n, m}  \right\rVert^2,
\end{equation}
where
\begin{equation}
E_{n, m} \doteq e^{j 2 \pi v_{n} t} e^{j 2 \pi \tau_{n} f} e^{j 2 \pi v_{m} t} e^{j 2 \pi \tau_{m} f}.
\end{equation}
Additionally, the path loss is computed as in~\eqref{eq:pl}. Since the useful signal received at D experiences two channels (from S to R, and from R to D) as a cascade, as described in~\eqref{eq:mimo_input_output_relay}, two path loss terms are added (in dB), to obtain the final \gls{psd} of $x_S$ at D as:
\begin{equation}
\begin{split}
\mathcal{P}_{r x}(t, f)& [\mathrm{dB}]  = \text{PL} (d_{\mathrm{SR}}, f_c) [\mathrm{dB}] \\
& + \text{PL} (d_{\mathrm{RD}}, f_c) [\mathrm{dB}] + \mathcal{\tilde{P}}_{rx}(t, f) [\mathrm{dB}].
\end{split}
\end{equation}
\item \emph{Interference and \gls{sinr}.}
As the last step, we evaluate the \glspl{psd} $\{ \mathcal{P}_{i}(t, f) \}_{i = 1, \dots, N_I}$ of the $N_I$ interfering signals at D. 
To do so, we follow Steps 1--4 as for the useful component of the signal. However, the beamforming configurations are not optimized as described in Step 2. That is to say, each interferer uses the beamforming vector yielding the highest \gls{sinr} \emph{towards its intended destination}, while R and D employ the same configurations used in the previous steps.
Finally, the \gls{sinr} is evaluated as:
\[ \Lambda (t, f) = \frac{ \mathcal{P}_{r x}(t, f)} {\sum_{i=1}^{N_I} \mathcal{P}_{i}(t, f) + \mathcal{P}_{n}(t, f) },  \]
where $\mathcal{P}_{n}(t, f)$ is the \gls{psd} of the thermal noise at D.
\end{enumerate}

\subsection{Integration of the IRS/AF Signal Model in the Simulator}
In Sec.~\ref{sec:ext_ch_model} we described how our simulator computes the channel (in terms of \gls{psd}) in case of IRS/AF relays, which is then used to calculate the end-to-end SINR at the destination D. 
Notice that the \gls{sinr} can refer to either the \gls{sinr} relative to the whole bandwidth, for narrowband signals over frequency-flat channels, or the \gls{sinr} experienced over a single subcarrier, for wideband signals transmitted over frequency-selective channels. 
In the second case, the \glspl{sinr} corresponding to the various frequency chunks are then mapped into a single \gls{sinr} value, according to additional maps obtained from link-level simulations~\cite{lagen2020new}.
Based on that, our simulator defines a \gls{l2sm}, i.e., a table which associates a given \gls{sinr} to a \gls{mac}-layer \gls{tb} error rate~\cite{mezzavilla2012lightweight}, in turn used to  decide whether the \gls{tb} has been correctly received or not.

The upper layers of the 5G NR protocol stack are modeled based on the \texttt{ns3-mmwave} module~\cite{mezzavilla2018end}.
It implements a custom \gls{phy} layer supporting the NR frame structures and numerologies, and a \gls{mac} layer with ad hoc beamforming and scheduling policies. 
The \gls{rlc} and \gls{pdcp} layers implement network functions such as packet segmentation, retransmissions and/or reassembly. 



\section{Performance Evaluation}
\label{sec:results}

In this section we describe our simulation setup and parameters (Sec.~\ref{sub:simulation_setup}), and evaluate the performance of \glspl{irs} and \gls{af} relays, considering full-stack network metrics as a function of different antenna array configurations (Sec.~\ref{sub:numerical_results}).

\subsection{Simulation Setup} 
\label{sub:simulation_setup}

\begin{figure}[t!]
  \centering
    \begin{subfigure}[t]{0.8\columnwidth}
    \includegraphics[width=1\columnwidth]{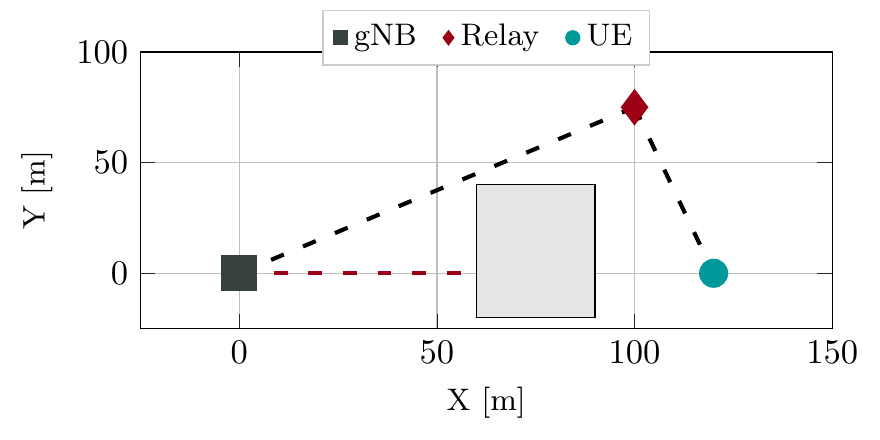}
    \caption{Scenario 1, with $N_{\rm U} = 1$.}
    \label{Fig:s1}
    \end{subfigure}
     \begin{subfigure}[t]{0.8\columnwidth}
    \includegraphics[width=1\columnwidth]{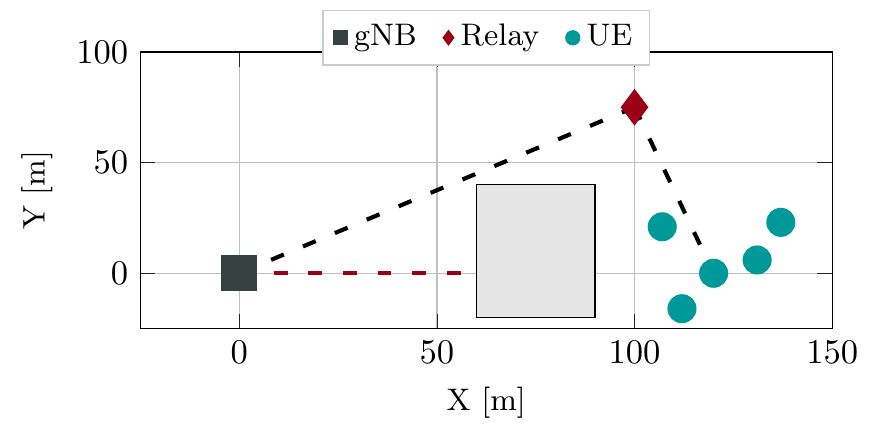}
    \caption{Scenario 2, with $N_{\rm U} = 5$.}
    \label{Fig:s2}
    \end{subfigure}
     \caption{Simulation scenarios, where we deploy one \gls{gnb}, $N_{\rm U}$ \glspl{ue} and, possibly, a relay. A building (the gray rectangle) blocks the direct link (dashed red line) from the \gls{gnb} to the \glspl{ue}. In turn, the relay guarantees a \gls{los} link (dashed black line) to all the devices.}
    \label{Fig:scenarios}
\end{figure}

In our simulations we consider two simple yet realistic urban canyon scenarios, where we deploy a single \gls{gnb}, $N_{\rm U}$ \glspl{ue}, with $N_{\rm U}=1$ (5) in Scenario 1 (2), as illustrated in Fig.~\ref{Fig:scenarios}, and a single relay, which can be either an \gls{irs} or an \gls{af} relay.
The wireless channel is modeled as an \gls{uma} link~\cite{3gpp.38.901}. 
The \gls{los}/\gls{nlos} condition depends on the geometry of the scenario.
In particular, we assume that the direct wireless link between the \glspl{ue} and the \gls{gnb} is blocked by a building, as illustrated in Fig.~\ref{Fig:scenarios}, which introduces an additional penetration loss modeled based on~\cite[Sec. 7.4.3.1]{3gpp.38.901}.
The end nodes can still communicate in \gls{los} via the relay. 
 


Our simulation parameters are reported in Table~\ref{Tab:parameters}.
Specifically, the \glspl{ue} download \gls{udp} data, modeled as a constant bit-rate stream of 50~Mbps, from a remote server. 
We assume that, at each transmission opportunity towards the generic $k$-th \gls{ue}, both \gls{af} and \gls{irs} relays can use their optimal configuration, i.e., the codeword yielding the highest end-to-end \gls{sinr} towards \gls{ue} $k$.
The system operates at 28 GHz, with a total bandwidth of 100~MHz, to be shared among all the devices in \gls{tdma}. 
The gNB is equipped with an antenna array of 64 elements, and uses a power of 33 dBm.
For the IRS, we consider a number of reflecting elements from 200 to 7\,200.
For the AF relay, we consider antenna arrays from 16 to 256 elements.


\def\arraystretch{1.3}
\begin{table}[t!]
  \caption{Simulation parameters.}
  \label{Tab:parameters}
  \centering
  \scriptsize
  \begin{tabular}{l|l}
    \toprule
    Parameter                  & Value \\ \midrule
    Carrier frequency		   & 28~GHz	\\
    Total bandwidth			   & 100~MHz \\
    Number of \glspl{ue} ($N_{\rm U}$)  & \{1, 5\} \\
    \gls{gnb} antenna array    & $8$H$\times8$V \\
    \gls{gnb} max RF power	   & 33~dBm \\
    \gls{ue} antenna array     & $2$H$\times1$V \\
    \gls{irs} antenna array    & \{$10$H$\times20$V, $20$H$\times40$V,$ 40$H$\times80$V, $60$H$\times120$V\} \\
    \gls{af} antenna array     & \{$4$H $\times4$V, $8$H $\times8$V, $16$H $\times16$V\} \\
    \gls{af} amplification & 40~dB \\
    Antenna radiation pattern  & \cite[Table 7.3-1]{3gpp.38.901} \\
    UDP source rate			   & 50~Mbps\\
    \bottomrule
  \end{tabular}
\end{table}

\subsection{Numerical Results} 
\label{sub:numerical_results}

We now compare the end-to-end performance of \gls{irs}- and \gls{af}-relay assisted networks in terms of:
\begin{itemize}
    \item \emph{\gls{sinr}}. It is a measure of the quality of the channel. 
    It depends on PHY-layer characteristics, including the relative distance between the transmitter, the receiver and the relay (if applicable), the operating frequency, the propagation conditions, 
    and the channel bandwidth. 
    \item \emph{End-to-end throughput}. 
     It is measured as the total number of received bytes per user divided by the total simulation time. 
    \item \emph{End-to-end latency.} It is measured from the time each packet is generated at the application layer to when it is successfully received.
  Accordingly, it accounts for both transmission and queuing times. 
 \item \emph{\gls{per}}. It is measured as the ratio between the number of packets delivered with errors and the total number of transmitted packets.
\end{itemize}
The IRS/AF performance will be evaluated against a baseline scenario (referred to as “gNB-only”) in which there is no intermediate relay.

\begin{figure}
  \centering
  \begin{subfigure}[t]{\columnwidth}
    \centering
    \setlength\fwidth{0.8\columnwidth}
    \setlength\fheight{0.5\columnwidth}
    \input{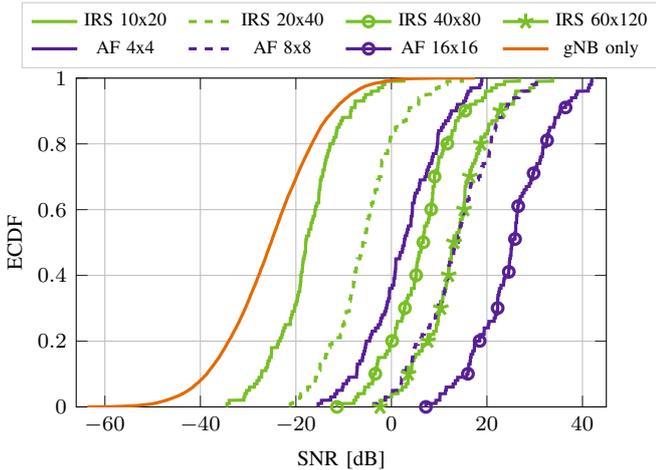}
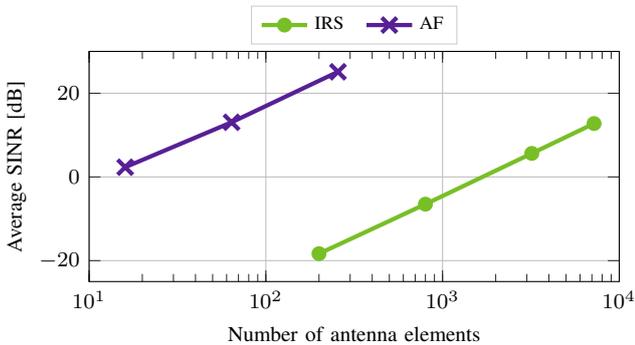
    \vspace*{-3mm}
    \caption{ECDF of the \gls{sinr} for different relay configurations.\vspace{0.5cm}}
    \label{Fig:sinr_all}
  \end{subfigure}
 \hfill
  \begin{subfigure}[t]{\columnwidth}
    \centering
    \setlength\fwidth{0.8\columnwidth}
    \setlength\fheight{0.35\columnwidth}
    \input{Figures/Results/1UE/Avg_sinr_vs_N_1ue.tex}
    \vspace*{-3mm}
    \caption{Average \gls{sinr} vs. the number of radiating elements at the relay.}
    \vspace*{3mm}
    \label{Fig:sinr_vs_n}
  \end{subfigure}
  \caption{\gls{sinr} statistics for Scenario~1.}
  \label{Fig:sinr}
\end{figure}

\paragraph{SINR} Our analysis starts with the \gls{sinr} statistics depicted in Fig.~\ref{Fig:sinr}, relative to Scenario 1 with $N_{\rm U}=1$.
First, in Fig.~\ref{Fig:sinr_all} we observe that the presence of the relay improves the \gls{sinr} (on average up to $+55$ dB) compared to the ``gNB only'' baseline, in which the \gls{ue} communicates in \gls{nlos}. 
Notably, as depicted in Fig.~\ref{Fig:sinr_vs_n}, both \gls{irs} and \gls{af} relays provide an end-to-end \gls{sinr} gain which scales proportionally 
with respect to the number of radiating elements at the relay. 
For the \gls{irs}, this effect is given by the beamforming gain, as well as by the fact that the power collected by the \gls{irs} is proportional to its surface area, which in turn is proportional to the number of radiating elements~\cite{bjornson2020reconfigurable}.

The AF-assisted configurations always outperform the IRS-assisted ones in terms of SINR (on average up to $+40$ dB, with the same number of antennas): this is expected since the AF relay amplifies the signal, thus achieving a higher end-to-end gain. 
Notice that the SINR is below 0 dB when the IRS is made of less that $800$ elements, which justifies the use of very large IRS panels. 
Indeed, an \gls{irs} panel of $60\times120$ elements provides an average SINR of $13$ dB, which is enough to support reliable transmissions as long as communication requirements are not too extreme, as we will demonstrate in the following paragraphs. 


\begin{figure}
	\centering
	\setlength\fwidth{0.8\columnwidth}
	\setlength\fheight{0.5\columnwidth}
	\input{Figures/Results/Joined/E2E_Throughput_avg.tex}
	\caption{End-to-end per-\gls{ue} throughput at the application layer in Scenario~1 (wide bars) and Scenario 2 (narrow bars) for different relay configurations.}
	\label{fig:throughput}
\end{figure}
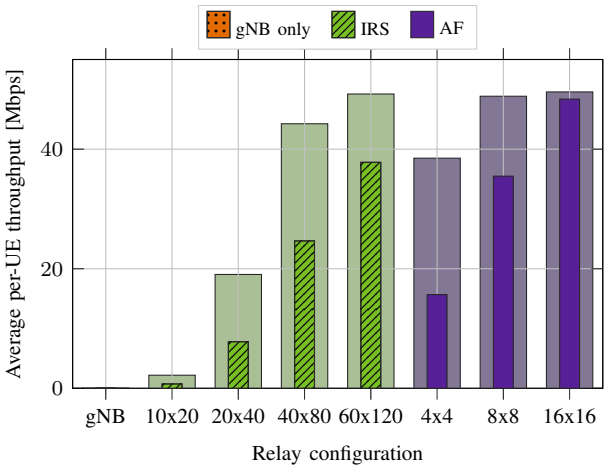

\paragraph{End-to-end throughput}
In Fig.~\ref{fig:throughput} we plot the end-to-end throughput experienced at the application layer, thus considering the impact of the whole 5G NR protocol stack.
When $N_{\rm U}=1$ (Scenario 1) the average throughput is an indication of the ergodic capacity.
We see that the throughput for the ``gNB only'' baseline is zero, given the very low SINR (below the sensitivity threshold of most commercial receivers) experienced at the physical layer. 
Interestingly, even though the AF relay with $16\times 16$ antennas guarantees, on average, 15 dB higher SINR than an IRS with $60\times120$ elements (from Fig.~\ref{Fig:sinr_all}), we see that the end-to-end throughput of the two configurations is comparable. 
This demonstrates that, in a simple scenario with only one UE, an average SINR of 15 dB is enough to satisfy all traffic requests. 
In this case, the IRS is more desirable than an AF relay given its simplicity. Also, it is not convenient to further increase the IRS size, given that the throughput is already maximized and equal to the UDP source rate (50 Mbps in our simulations).

When $N_{\rm U}=5$ (Scenario 2) the average per-\gls{ue} throughput decreases significantly with respect to Scenario 1 due to the fact that, in a multi-user scenario, radio resources must be shared among \glspl{ue}, which may lead to channel congestion. This result validates the accuracy and realism of our ns-3 framework.
 Nevertheless, this effect is less pronounced for very large antenna panels. For example, for an AF relay of $4\times4$ antennas, the per-\gls{ue} throughput drops by almost $60\%$, while considering an array of $16\times16$ elements the per-\gls{ue} throughput decreases by only $2\%$.
Even in Scenario 2, AF-assisted networks can still sustain the application source rate, as long as at least $16\times16$ antennas are used.
On the other hand, IRSs are constrained by the limited SINR available at the \gls{phy} layer, and are never able to achieve the full source rate offered by the application. The maximum achievable throughput is around $40$ Mbps for $60\times120$ elements, i.e., $-20$\% compared to the case of $N_{\rm U}=1$.



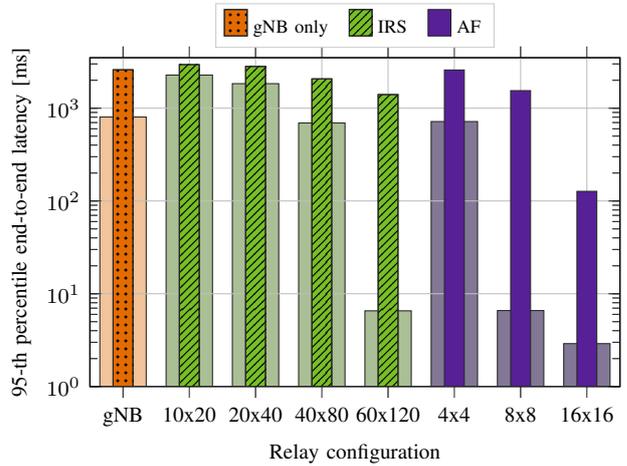
\begin{figure}
	\centering
	\setlength\fwidth{0.8\columnwidth}
	\setlength\fheight{0.5\columnwidth}
	\input{Figures/Results/Joined/E2E_Latency_95th.tex}
	\caption{95-th percentile of the end-to-end latency at the application layer in Scenario~1 (wide bars) and Scenario 2 (narrow bars) for different relay configurations.}
	\label{fig:latency}
\end{figure}

\begin{figure}
	\centering
	\setlength\fwidth{0.8\columnwidth}
	\setlength\fheight{0.5\columnwidth}
	\input{Figures/Results/Joined/PER.tex}
	\caption{Average PER at the application layer in Scenario~1 (wide bars) and Scenario 2 (narrow bars) for different relay configurations.}
	\label{fig:per}
\end{figure}
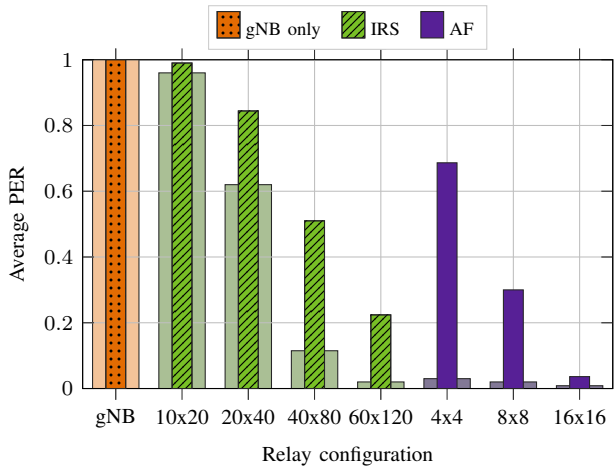

\paragraph{End-to-end latency}

Finally, in Fig.~\ref{fig:latency} we plot the 95-th percentile of the end-to-end latency experienced at the application layer.
 We can see that the performance is generally poor even in the simple scenario in which only one UE is deployed (Scenario 1), where the latency is higher than $100$~ms for most relay configurations, suggesting that in these cases the system is unstable.  In fact, the use of relays featuring small antenna panels results in very high levels of queuing and buffering, which leads to latency degradation. This issue can be solved by configuring larger IRS and AF relays, despite the increased system complexity. 
 For example, an IRS of $60\times120$ elements and an AF relay with $\geq 8\times8$ elements can guarantee an end-to-end latency lower than $10$~ms, that is in line with most 5G application requirements. 
 Notice that the latency for the ``\gls{gnb} only'' configuration is not particularly representative, as it is relative to only the correctly received packets.
 In fact, without the relay, transmissions are in \gls{nlos} and result in several packet losses (see the \gls{per} in Fig.~\ref{fig:per}), which makes the system less congested; the (few) packets that make it to the application layer are then transmitted with lower delay. Nevertheless, the latency is still more than two orders of magnitude higher than considering the best IRS and AF configurations, an indication that relays are desirable in these types of networks.


When $N_{\rm U}=5$ (Scenario 2), the latency is generally higher compared to when $N_{\rm U}=1$. This is expected since \glspl{ue} are competing for the available resources. In addition, using \gls{udp} as transport protocol, thus with a full buffer source traffic model, each end-to-end flow does not self-regulate to the actual network conditions, thus congestion arises. 
Better performance could be achieved considering non-UDP traffic: for example, the congestion control mechanism available in \gls{tcp} would regulate the source traffic, and prevent network congestion and buffer~overflow.

Notice that, even considering the most aggressive IRS architecture with $60\times120$ elements, the latency is on average above $1000$ ms, vs. $6.5$ ms in Scenario 1. 
This is due to the fact that, in this scenario, more than 20\% of the packets are lost and retransmitted (see Fig.~\ref{fig:per}), which increases the packet delay.
For an AF relay with $16\times16$ antennas, instead, the latency is more than $10$ times lower and equal to around $130$~ms on average, with a \gls{per} as low as 3\%, which can still support some key target communication requirements.
We can conclude that IRS-assisted networks, despite consuming less power, are not appropriate in this scenario, unless very large IRS panels are~used.

\section{Conclusions and Future Work}
\label{sec:conclusion}
\Glspl{irs} and \gls{af} relays are amongst the most promising technologies to facilitate 6G networks.
Not only can these elements improve both communication and coverage of wireless devices, but also promote lower energy consumption compared to IAB systems.
In this paper we proposed a signal model for \glspl{irs} and \gls{af} relays, based on the 3GPP TR 38.901 channel for 5G NR networks, and explained the methodology we used to perform network-level simulations of 5G scenarios with IRS and AF relay nodes.
Based on this framework, we performed simulations to provide numerical guidelines to dimension IRS/AF-assisted networks. In particular we obtained that:
\begin{itemize}
	\item Both IRS and AF relays can improve the throughput, latency and PER of end users compared to a baseline scenario in which relays are not deployed.
	\item IRSs are valid solutions in small networks, and more desirable technologies than AF relays given their inherent simplicity and power efficiency.
	\item AF relay are more appropriate in dense networks, while IRSs should be large, despite the increased system complexity, to satisfy the typical communication requirements.
\end{itemize}

As part of our future research we will further extend our network simulator to consider more sophisticated/advanced scenarios, for example in which heterogeneous types of relays are deployed, and compare the numerical performance of IRS/AF relays with that of IAB.

\section*{Acknowledgment} 
The work of Amir Ashtari Gargari received support by the European Commission through the EU MSCA ITN Program (grant no. 861222). 

\bibliographystyle{IEEEtran}
\bibliography{bibl.bib}

\end{document}

%% file: acronymsSorted.tex
\newacronym{ca}{CA}{Carrier Aggregation}
\newacronym{3gpp}{3GPP}{3rd Generation Partnership Project}
\newacronym{5g}{5G}{5th generation}
\newacronym{5gc}{5GC}{5G Core}
\newacronym{adc}{ADC}{Analog to Digital Converter}
\newacronym{afbw}{AFBW}{Average Fading Bandwidth}
\newacronym{aimd}{AIMD}{Additive Increase Multiplicative Decrease}
\newacronym{am}{AM}{Acknowledged Mode}
\newacronym{amc}{AMC}{Adaptive Modulation and Coding}
\newacronym{aoa}{AoA}{Angle of Arrival}
\newacronym{aod}{AoD}{Angle of Departure}
\newacronym{ap}{AP}{Access Point}
\newacronym{af}{AF}{Amplify-and-Forward}
\newacronym{rr}{RR}{regenerative relay}
\newacronym{aqm}{AQM}{Active Queue Management}
\newacronym{awgn}{AGWN}{Additive White Gaussian Noise}
\newacronym{balia}{BALIA}{Balanced Link Adaptation}
\newacronym{bdp}{BDP}{Bandwidth-Delay Product}
\newacronym{ber}{BER}{Bit Error Rate}
\newacronym{bf}{BF}{Beamforming}
\newacronym{bwp}{BWP}{Bandwidth Part}
\newacronym{cad}{CAD}{Computer-Aided Design}
\newacronym{cc}{CC}{Congestion Control}
\newacronym{cdf}{CDF}{Cumulative Distribution Function}
\newacronym{cir}{CIR}{Channel Impulse Response}
\newacronym{cn}{CN}{Core Network}
\newacronym{cp}{CP}{Control Plane}
\newacronym{cqi}{CQI}{Channel Quality Information}
\newacronym{crs}{CRS}{Cell Reference Signal}
\newacronym{csirs}{CSI-RS}{Channel State Information - Reference Signal}
\newacronym{dc}{DC}{Dual Connectivity}
\newacronym{dce}{DCE}{Direct Code Execution}
\newacronym{dci}{DCI}{Downlink Control Information}
\newacronym{dl}{DL}{Downlink}
\newacronym{dmr}{DMR}{Deadline Miss Ratio}
\newacronym{dmrs}{DMRS}{DeModulation Reference Signal}
\newacronym{dray}{D-Ray}{Deterministic Ray}
\newacronym{e2e}{E2E}{end-to-end}
\newacronym{ecn}{ECN}{Explicit Congestion Notification}
\newacronym{edf}{EDF}{Earliest Deadline First}
\newacronym{em}{EM}{electromagnetic}
\newacronym{enb}{eNB}{evolved Node Base}
\newacronym{endc}{EN-DC}{E-UTRAN-\gls{nr} \gls{dc}}
\newacronym{epc}{EPC}{Evolved Packet Core}
\newacronym{es}{ES}{Edge Server}
\newacronym{fdd}{FDD}{Frequency Division Duplexing}
\newacronym{fdma}{FDMA}{Frequency Division Multiple Access}
\newacronym{fray}{F-Ray}{Flashing Ray}
\newacronym{fs}{FS}{Fast Switching}
\newacronym{ftp}{FTP}{File Transfer Protocol}
\newacronym{gmm}{GMM}{Gaussian Mixture Model}
\newacronym{gnb}{gNB}{Next Generation Node Base}
\newacronym{harq}{HARQ}{Hybrid Automatic Repeat reQuest}
\newacronym{hetnet}{HetNet}{Heterogeneous Network}
\newacronym{hh}{HH}{Hard Handover}
\newacronym{hol}{HOL}{Head-of-Line}
\newacronym{hqf}{HQF}{Highest-quality-first}
\newacronym{ia}{IA}{Initial Access}
\newacronym{iab}{IAB}{Integrated Access and Backhaul}
\newacronym{imt}{IMT}{International Mobile Telecommunication}
\newacronym{inf}{InF}{Indoor Factory}
\newacronym{irs}{IRS}{Intelligent Reflective Surface}
\newacronym{inr}{INR}{Interference to Noise Ratio}
\newacronym{iot}{IoT}{Internet of Things}
\newacronym{ked}{KED}{Knife-Edge Diffraction}
\newacronym{kpi}{KPI}{Key Performance Indicator}
\newacronym{lcf}{LCF}{Level Crossing Frequency}
\newacronym{lcr}{LCR}{Level Crossing Rate}
\newacronym{los}{LoS}{Line-of-Sight}
\newacronym{l2sm}{L2SM}{Link-to-System Mapping}
\newacronym{lte}{LTE}{Long Term Evolution}
\newacronym{ltemtp}{LTE-M}{LTE-MTC [Machine Type Communication]}
\newacronym{m2m}{M2M}{Machine to Machine}
\newacronym{mac}{MAC}{Medium Access Control}
\newacronym{mc}{MC}{Monte Carlo}
\newacronym{mcs}{MCS}{Modulation and Coding Scheme}
\newacronym{mec}{MEC}{Mobile Edge Cloud}
\newacronym{mi}{MI}{Mutual Information}
\newacronym{mib}{MIB}{Master Information Block}
\newacronym{mimo}{MIMO}{Multiple Input Multiple Output}
\newacronym{m-mimo}{m-MIMO}{massive-MIMO}
\newacronym{mlr}{MLR}{Maximum-local-rate}
\newacronym{mmwave}{mmWave}{millimeter wave}
\newacronym{moi}{MoI}{Method of Images}
\newacronym{mpc}{MPC}{Multi Path Component}
\newacronym{mptcp}{MPTCP}{Multipath TCP}
\newacronym{mr}{MR}{Maximum Rate}
\newacronym{mrdc}{MR-DC}{Multi \gls{rat} \gls{dc}}
\newacronym{mss}{MSS}{Maximum Segment Size}
\newacronym{mtd}{MTD}{Machine-Type Device}
\newacronym{mtu}{MTU}{Maximum Transmission Unit}
\newacronym{nfv}{NFV}{Network Function Virtualization}
\newacronym{nist}{NIST}{National Institute of Standards and Technology}
\newacronym{nlos}{NLoS}{Non-Line-of-Sight}
\newacronym{ntn}{NTN}{Non-Terrestrial Networks}
\newacronym{nr}{NR}{New Radio}
\newacronym{nrmse}{NRMSE}{Normalized Root Mean Square Error}
\newacronym{nsa}{NSA}{Non Stand Alone}
\newacronym{o2i}{O2I}{Outdoor-to-Indoor}
\newacronym{ofdm}{OFDM}{Orthogonal Frequency Division Multiplexing}
\newacronym{pa}{PA}{power amplifier}
\newacronym{prr}{PRR}{Packet Reception Ratio}
\newacronym{pbch}{PBCH}{Physical Broadcast Channel}
\newacronym{pdcch}{PDCCH}{Physical Downlonk Control Channel}
\newacronym{pdcp}{PDCP}{Packet Data Convergence Protocol}
\newacronym{pdsch}{PDSCH}{Physical Downlink Shared Channel}
\newacronym{pdu}{PDU}{Packet Data Unit}
\newacronym{per}{PER}{Packet Error Rate}
\newacronym{pf}{PF}{Proportional Fair}
\newacronym{pgw}{PGW}{Packet Gateway}
\newacronym{phy}{PHY}{Physical}
\newacronym{pl}{PL}{Path Loss}
\newacronym{ppp}{PPP}{Poisson Point Process}
\newacronym{prb}{PRB}{Physical Resource Block}
\newacronym{psd}{PSD}{Power Spectral Density}
\newacronym{pss}{PSS}{Primary Synchronization Signal}
\newacronym{pucch}{PUCCH}{Physical Uplink Control Channel}
\newacronym{pusch}{PUSCH}{Physical Uplink Shared Channel}
\newacronym{qam}{QAM}{Quadrature Amplitude Modulation}
\newacronym{qd}{QD}{Quasi Deterministic}
\newacronym{rach}{RACH}{Random Access Channel}
\newacronym{ran}{RAN}{Radio Access Network}
\newacronym[firstplural=Radio Access Technologies (RATs)]{rat}{RAT}{Radio Access Technology}
\newacronym{red}{RED}{Random Early Detection}
\newacronym{rf}{RF}{Radio Frequency}
\newacronym{rlc}{RLC}{Radio Link Control}
\newacronym{rlf}{RLF}{Radio Link Failure}
\newacronym{rray}{R-Ray}{Random Ray}
\newacronym{rrc}{RRC}{Radio Resource Control}
\newacronym{rrm}{RRM}{Radio Resource Management}
\newacronym{rs}{RS}{Remote Server}
\newacronym{rsrp}{RSRP}{Reference Signal Received Power}
\newacronym{rsrq}{RSRQ}{Reference Signal Received Quality}
\newacronym{rss}{RSS}{Received Signal Strength}
\newacronym{rssi}{RSSI}{Received Signal Strength Indicator}
\newacronym{rt}{RT}{Ray Tracer}
\newacronym{rtt}{RTT}{Round Trip Time}
\newacronym{rw}{RW}{Receive Window}
\newacronym{rx}{RX}{Receiver}
\newacronym{sa}{SA}{standalone}
\newacronym{sack}{SACK}{Selective Acknowledgment}
\newacronym{sap}{SAP}{Service Access Point}
\newacronym{sch}{SCH}{Secondary Cell Handover}
\newacronym{scm}{SCM}{Spatial Channel Model}
\newacronym{scoot}{SCOOT}{Split Cycle Offset Optimization Technique}
\newacronym{sdap}{SDAP}{Service Data Adaptation Protocol}
\newacronym{sdma}{SDMA}{Spatial Division Multiple Access}
\newacronym{sf}{SF}{Shadow Fading}
\newacronym{si}{SI}{Study Item}
\newacronym{sib}{SIB}{Secondary Information Block}
\newacronym{sinr}{SINR}{Signal-to-Interference-plus-Noise Ratio}
\newacronym{sir}{SIR}{Signal-to-Interference Ratio}
\newacronym{sm}{SM}{Saturation Mode}
\newacronym{snr}{SNR}{Signal-to-Noise Ratio}
\newacronym{son}{SON}{Self-Organizing Network}
\newacronym{srs}{SRS}{Sounding Reference Signal}
\newacronym{ss}{SS}{Synchronization Signal}
\newacronym{sss}{SSS}{Secondary Synchronization Signal}
\newacronym{sta}{STA}{Station}
\newacronym{tb}{TB}{Transport Block}
\newacronym{tcp}{TCP}{Transmission Control Protocol}
\newacronym{udp}{UDP}{User Datagram Protocol}
\newacronym{tdd}{TDD}{Time Division Duplexing}
\newacronym{tdma}{TDMA}{Time Division Multiple Access}
\newacronym{tfl}{TfL}{Transport for London}
\newacronym{tgad}{TGad}{Task Group ad}
\newacronym{tgay}{TGay}{Task Group ay}
\newacronym{tm}{TM}{Transparent Mode}
\newacronym{trp}{TRP}{Transmitter Receiver Pair}
\newacronym{tti}{TTI}{Transmission Time Interval}
\newacronym{ttt}{TTT}{Time-to-Trigger}
\newacronym{tx}{TX}{Transmitter}
\newacronym{ue}{UE}{User Equipment}
\newacronym{ul}{UL}{Uplink}
\newacronym{um}{UM}{Unacknowledged Mode}
\newacronym{uma}{UMa}{Urban Macro}
\newacronym{uml}{UML}{Unified Modeling Language}
\newacronym{utc}{UTC}{Urban Traffic Control}
\newacronym{vm}{VM}{Virtual Machine}
\newacronym{wbf}{WBF}{Wired Bias Function}
\newacronym{wf}{WF}{Wired-first}
\newacronym{wifi}{Wi-Fi}{Wireless Fidelity}
\newacronym{wigig}{WiGig}{Wireless Gigabit}
\newacronym{wlan}{WLAN}{Wireless Local Area Network}
\newacronym{xpr}{XPR}{Cross Polarization Ratio}
\newacronym{fr2}{FR2}{Frequency Range 2}
\newacronym{nbiot}{NB-IoT}{Narrowband-IoT}
\newacronym{cps}{CPS}{Cyber-Physical production System}
\newacronym{iiot}{IIoT}{Industrial Internet of Things}
\newacronym{agv}{AGV}{Autonomous Ground Vehicle}
\newacronym{uav}{UAV}{Unmanned Autonomous Vehicle}
\newacronym{amr}{AMR}{Autonomous Mobile Robots}
\newacronym{wsn}{WSN}{Wireless Sensor Network}
\newacronym{embb}{eMBB}{enhanced Mobile Broadband}
\newacronym{urllc}{URLLC}{Ultra-Reliable Low-Latency Communications}

%% file: myTikz.tex
\tikzstyle{startstop} = [rectangle, rounded corners, minimum width=2cm, minimum height=0.5cm,text centered, draw=black]
\tikzstyle{io} = [trapezium, trapezium left angle=70, trapezium right angle=110, minimum width=3cm, minimum height=1cm, text centered, draw=black]
\tikzstyle{process} = [rectangle, minimum width=2cm, minimum height=0.5cm, text centered, draw=black, alignb=center]
\tikzstyle{decision} = [ellipse, minimum width=2cm, minimum height=1cm, text centered, draw=black]
\tikzstyle{arrow} = [thick,<->,>=stealth]
\tikzstyle{line} = [thick,>=stealth]
\tikzstyle{darrow} = [thick,<->,>=stealth,dashed]
\tikzstyle{sarrow} = [thick,->,>=stealth]
\tikzstyle{larrow} = [line width=0.1mm,dashdotted,->,>=stealth]

\makeatletter
\def\grd@save@target#1{%
  \def\grd@target{#1}}
\def\grd@save@start#1{%
  \def\grd@start{#1}}
\tikzset{
  grid with coordinates/.style={
    to path={%
      \pgfextra{%
        \edef\grd@@target{(\tikztotarget)}%
        \tikz@scan@one@point\grd@save@target\grd@@target\relax
        \edef\grd@@start{(\tikztostart)}%
        \tikz@scan@one@point\grd@save@start\grd@@start\relax
        \draw[minor help lines] (\tikztostart) grid (\tikztotarget);
        \draw[major help lines] (\tikztostart) grid (\tikztotarget);
        \grd@start
        \pgfmathsetmacro{\grd@xa}{\the\pgf@x/1cm}
        \pgfmathsetmacro{\grd@ya}{\the\pgf@y/1cm}
        \grd@target
        \pgfmathsetmacro{\grd@xb}{\the\pgf@x/1cm}
        \pgfmathsetmacro{\grd@yb}{\the\pgf@y/1cm}
        \pgfmathsetmacro{\grd@xc}{\grd@xa + \pgfkeysvalueof{/tikz/grid with coordinates/major step x}}
        \pgfmathsetmacro{\grd@yc}{\grd@ya + \pgfkeysvalueof{/tikz/grid with coordinates/major step y}}
        \foreach \x in {\grd@xa,\grd@xc,...,\grd@xb}
        \node[anchor=north] at (\x,\grd@ya) {\pgfmathprintnumber{\x}};
        \foreach \y in {\grd@ya,\grd@yc,...,\grd@yb}
        \node[anchor=east] at (\grd@xa,\y) {\pgfmathprintnumber{\y}};
      }
    }
  },
  minor help lines/.style={
    help lines,
    gray,
    line cap =round,
    xstep=\pgfkeysvalueof{/tikz/grid with coordinates/minor step x},
    ystep=\pgfkeysvalueof{/tikz/grid with coordinates/minor step y}
  },
  major help lines/.style={
    help lines,
    line cap =round,
    line width=\pgfkeysvalueof{/tikz/grid with coordinates/major line width},
    xstep=\pgfkeysvalueof{/tikz/grid with coordinates/major step x},
    ystep=\pgfkeysvalueof{/tikz/grid with coordinates/major step y}
  },
  grid with coordinates/.cd,
  minor step x/.initial=.5,
  minor step y/.initial=.2,
  major step x/.initial=1,
  major step y/.initial=1,
  major line width/.initial=1pt,
}
\makeatother

%% file: Figures/Results/1UE/Avg_sinr_vs_N_1ue.tex
\begin{tikzpicture}

  \pgfplotsset{every tick label/.append style={font=\footnotesize}}
  \definecolor{plotColor1}{HTML}{E36B02}
  \definecolor{plotColor2}{HTML}{78BF26}
  \definecolor{plotColor3}{HTML}{572096}
  \definecolor{plotColor4}{HTML}{338EDE}
  \definecolor{plotColor1Light}{HTML}{F3C399}
  \definecolor{plotColor2Light}{HTML}{AABF95}
  \definecolor{plotColor3Light}{HTML}{827896}
  \definecolor{plotColor4Light}{HTML}{B2C8DE}
  \definecolor{color6}{rgb}{0.890196078431372,0.466666666666667,0.76078431372549}

  \begin{semilogxaxis}[
    width=\fwidth,
    height=\fheight,
    at={(0\fwidth,0\fheight)},
    scale only axis,
    legend image post style={mark indices={}},
    legend style={
        /tikz/every even column/.append style={column sep=0.2cm},
        at={(0.5, 1.03)}, 
        anchor=south, 
        draw=white!80!black, 
        font=\scriptsize
        },
    legend columns=3,
    xlabel style={font=\footnotesize},
    xlabel={Number of antenna elements},
    xmajorgrids,
    xmin=10, xmax=1e4,
    xtick style={color=white!15!black},
    ylabel shift = -2 pt,
    ylabel style={font=\footnotesize},
    ylabel={Average SINR [dB]},
    ymajorgrids,
    ymajorticks=true,
    ymin=-25, ymax=30,
    ytick style={color=white!15!black},
]
\addplot [ultra thick, plotColor2, mark=*]
table {%
200 -18.33762033
800 -6.506891
3200 5.6205
7200 12.763146246
};
\addlegendentry{IRS}
\addplot [ultra thick, plotColor3, mark=x, mark size=4]
table {%
16	2.325820
64	13.0615893
256	25.1253
};
\addlegendentry{AF}

\end{semilogxaxis}

\end{tikzpicture}

%% file: Figures/Results/Joined/E2E_Throughput_avg.tex
\begin{tikzpicture}

    \pgfplotsset{every tick label/.append style={font=\footnotesize}}
    \definecolor{plotColor1}{HTML}{E36B02}
    \definecolor{plotColor2}{HTML}{78BF26}
    \definecolor{plotColor3}{HTML}{572096}
    \definecolor{plotColor4}{HTML}{338EDE}
    \definecolor{plotColor1Light}{HTML}{F3C399}
    \definecolor{plotColor2Light}{HTML}{AABF95}
    \definecolor{plotColor3Light}{HTML}{827896}
    \definecolor{plotColor4Light}{HTML}{B2C8DE}
    \definecolor{color6}{rgb}{0.890196078431372,0.466666666666667,0.76078431372549}

\begin{axis}[
    width=\fwidth,
    height=\fheight,
    at={(0\fwidth,0\fheight)},
    scale only axis,
    legend image post style={mark indices={}},
    legend style={
        /tikz/every even column/.append style={column sep=0.2cm},
        at={(0.5, 1.03)}, 
        anchor=south, 
        draw=white!80!black, 
        font=\scriptsize
        },
    legend columns=3,
    xlabel style={font=\footnotesize},
    xlabel={},
    yticklabels={,,,,,,},
    xticklabels={,,,,,,},
    xmin=-0.5, xmax=7.5,
    ylabel shift = -1 pt,
    ylabel style={font=\footnotesize},
    ylabel={},
    ymin=0, ymax=55,
    ybar,
    bar width=.7,
    bar shift=0pt,
]

    \addplot[fill=plotColor1Light, draw=white!15!black] coordinates {(0, 0.02678)};
    \addplot[fill=plotColor2Light, draw=white!15!black] coordinates {(1, 2.18999)};
    \addplot[fill=plotColor2Light, draw=white!15!black, forget plot] coordinates {(2, 19.04253)};
    \addplot[fill=plotColor2Light, draw=white!15!black, forget plot] coordinates {(3, 44.25216)};
    \addplot[fill=plotColor2Light, draw=white!15!black, forget plot] coordinates {(4, 49.22552)};
    \addplot[fill=plotColor3Light, draw=white!15!black] coordinates {(5, 38.49637)};
    \addlegendentry{AF}
    \addplot[fill=plotColor3Light, draw=white!15!black] coordinates {(6, 48.85752)};
    \addplot[fill=plotColor3Light, draw=white!15!black] coordinates {(7, 49.575)};
\end{axis}

\begin{axis}[
    width=\fwidth,
    height=\fheight,
    at={(0\fwidth,0\fheight)},
    scale only axis,
    legend image post style={mark indices={}},
    legend style={
        /tikz/every even column/.append style={column sep=0.2cm},
        at={(0.5, 1.03)}, 
        anchor=south, 
        draw=white!80!black, 
        font=\scriptsize
        },
    legend columns=3,
    xlabel style={font=\footnotesize},
    xtick align=inside,
    ytick align=inside,
    xlabel={Relay configuration},
    xtick={0,1,2,3,4,5,6,7},
    xticklabels={gNB, 10x20, 20x40, 40x80, 60x120, 4x4, 8x8, 16x16},
    xmajorgrids,
    xmin=-0.5, xmax=7.5,
    xtick style={color=white!15!black},
    ylabel shift = -1 pt,
    ylabel style={font=\footnotesize, align=center},
    ylabel={Average per-UE throughput [Mbps]},
    ymajorgrids,
    ymin=0, ymax=55,
    ytick style={color=white!15!black},
    ybar,
    bar width=.3,
    bar shift=0pt,
    legend image code/.code={
        \draw [#1] (0cm,-0.1cm) rectangle (0.3cm, 0.25cm); }
]


    \addplot[fill=plotColor1, draw=white!15!black, postaction={pattern={dots}}] coordinates {(0, 0.00885)};
    \addlegendentry{gNB only}

    \addplot[fill=plotColor2, draw=white!15!black, postaction={pattern=north east lines}] 
        coordinates {(1, 0.7290)};
    \addlegendentry{IRS}
    \addplot[fill=plotColor2, draw=white!15!black, forget plot, postaction={pattern=north east lines}]  coordinates {(2, 7.790)};
    \addplot[fill=plotColor2, draw=white!15!black, forget plot, postaction={pattern=north east lines}]  coordinates {(3, 24.658)};
    \addplot[fill=plotColor2, draw=white!15!black, forget plot, postaction={pattern=north east lines}]  coordinates {(4, 37.800106)};
    \addplot[fill=plotColor3, draw=white!15!black] coordinates {(5, 15.679306666)};
    \addlegendentry{AF}
    \addplot[fill=plotColor3, draw=white!15!black] coordinates {(6, 35.486766666)};
    \addplot[fill=plotColor3, draw=white!15!black] coordinates {(7, 48.3738)};
    
\end{axis}

\end{tikzpicture}

%% file: Figures/Results/Joined/E2E_Latency_95th.tex
\begin{tikzpicture}

    \pgfplotsset{every tick label/.append style={font=\footnotesize}}
    \definecolor{plotColor1}{HTML}{E36B02}
    \definecolor{plotColor2}{HTML}{78BF26}
    \definecolor{plotColor3}{HTML}{572096}
    \definecolor{plotColor4}{HTML}{338EDE}
    \definecolor{plotColor1Light}{HTML}{F3C399}
    \definecolor{plotColor2Light}{HTML}{AABF95}
    \definecolor{plotColor3Light}{HTML}{827896}
    \definecolor{plotColor4Light}{HTML}{B2C8DE}
    \definecolor{color6}{rgb}{0.890196078431372,0.466666666666667,0.76078431372549}

\begin{semilogyaxis}[
    width=\fwidth,
    height=\fheight,
    at={(0\fwidth,0\fheight)},
    scale only axis,
    legend image post style={mark indices={}},
    legend style={
        /tikz/every even column/.append style={column sep=0.2cm},
        at={(0.5, 1.03)}, 
        anchor=south, 
        draw=white!80!black, 
        font=\scriptsize
        },
    legend columns=3,
    xlabel style={font=\footnotesize},
    xlabel={},
    yticklabels={,,,,,,},
    xticklabels={,,,,,,},
    xmin=-0.5, xmax=7.5,
    ylabel shift = -1 pt,
    ylabel style={font=\footnotesize},
    ylabel={},
    ymin=1, ymax=3500,
    ybar,
    bar width=.7,
    bar shift=0pt
]
    \addplot[fill=plotColor1Light, draw=white!15!black, forget plot] coordinates {(0, 803)};
    \addplot[fill=plotColor2Light, draw=white!15!black, forget plot] coordinates {(1, 2275)};
    \addplot[fill=plotColor2Light, draw=white!15!black, forget plot]  coordinates {(2, 1839)};
    \addplot[fill=plotColor2Light, draw=white!15!black, forget plot]  coordinates {(3, 693)};
    \addplot[fill=plotColor2Light, draw=white!15!black, forget plot]  coordinates {(4, 6.5471)};
    \addplot[fill=plotColor3Light, draw=white!15!black, forget plot] coordinates {(5, 717.974)};
    \addplot[fill=plotColor3Light, draw=white!15!black, forget plot] coordinates {(6, 6.589)};
    \addplot[fill=plotColor3Light, draw=white!15!black, forget plot] coordinates {(7, 2.9)};
\end{semilogyaxis}

\begin{semilogyaxis}[
    width=\fwidth,
    height=\fheight,
    at={(0\fwidth,0\fheight)},
    scale only axis,
    legend image post style={mark indices={}},
    legend style={
        /tikz/every even column/.append style={column sep=0.2cm},
        at={(0.5, 1.03)}, 
        anchor=south, 
        draw=white!80!black, 
        font=\scriptsize
        },
    legend columns=3,
    xlabel style={font=\footnotesize},
    xtick align=inside,
    ytick align=inside,
    xlabel={Relay configuration},
    xtick={0,1,2,3,4,5,6,7},
    xticklabels={gNB, 10x20, 20x40, 40x80, 60x120, 4x4, 8x8, 16x16},
    xmajorgrids,
    xmin=-0.5, xmax=7.5,
    xtick style={color=white!15!black},
    ylabel shift = -1 pt,
    ylabel style={font=\footnotesize, align=center},
    ylabel={95-th percentile end-to-end latency [ms]},
    ymajorgrids,
    ymin=1, ymax=3500,
    ytick style={color=white!15!black},
    ybar,
    bar width=.3,
    bar shift=0pt,
    legend image code/.code={
        \draw [#1] (0cm,-0.1cm) rectangle (0.3cm, 0.25cm); }
]
    \addplot[fill=plotColor1, draw=white!15!black, postaction={pattern={dots}}] coordinates {(0, 2596)};
    \addlegendentry{gNB only}
    \addplot[fill=plotColor2, draw=white!15!black, postaction={pattern=north east lines}] coordinates {(1, 2950)};
    \addlegendentry{IRS}
    \addplot[fill=plotColor2, draw=white!15!black, forget plot, postaction={pattern=north east lines}] coordinates {(2, 2817)};
    \addplot[fill=plotColor2, draw=white!15!black, forget plot, postaction={pattern=north east lines}] coordinates {(3, 2072)};
    \addplot[fill=plotColor2, draw=white!15!black, forget plot, postaction={pattern=north east lines}] coordinates {(4, 1402)};
    \addplot[fill=plotColor3, draw=white!15!black] coordinates {(5, 2580)};
    \addlegendentry{AF}
    \addplot[fill=plotColor3, draw=white!15!black] coordinates {(6, 1546)};
    \addplot[fill=plotColor3, draw=white!15!black] coordinates {(7, 127)};
\end{semilogyaxis}

\end{tikzpicture}

%% file: Figures/Results/Joined/PER.tex
\begin{tikzpicture}

    \pgfplotsset{every tick label/.append style={font=\footnotesize}}
    \definecolor{plotColor1}{HTML}{E36B02}
    \definecolor{plotColor2}{HTML}{78BF26}
    \definecolor{plotColor3}{HTML}{572096}
    \definecolor{plotColor4}{HTML}{338EDE}
    \definecolor{plotColor1Light}{HTML}{F3C399}
    \definecolor{plotColor2Light}{HTML}{AABF95}
    \definecolor{plotColor3Light}{HTML}{827896}
    \definecolor{plotColor4Light}{HTML}{B2C8DE}
    \definecolor{color6}{rgb}{0.890196078431372,0.466666666666667,0.76078431372549}

\begin{axis}[
    width=\fwidth,
    height=\fheight,
    at={(0\fwidth,0\fheight)},
    scale only axis,
    legend image post style={mark indices={}},
    legend style={
        /tikz/every even column/.append style={column sep=0.2cm},
        at={(0.5, 1.03)}, 
        anchor=south, 
        draw=white!80!black, 
        font=\scriptsize
        },
    legend columns=3,
    xlabel style={font=\footnotesize},
    xlabel={},
    yticklabels={,,,,,,},
    xticklabels={,,,,,,},
    xmin=-0.5, xmax=7.5,
    ylabel shift = -1 pt,
    ylabel style={font=\footnotesize},
    ylabel={},
    ymin=0, ymax=1,
    ybar,
    bar width=.7,
    bar shift=0pt,
]

    \addplot[fill=plotColor1Light, draw=white!15!black] coordinates {(0, 100)};
    \addplot[fill=plotColor2Light, draw=white!15!black] coordinates {(1, 0.96)};
    \addplot[fill=plotColor2Light, draw=white!15!black, forget plot] coordinates {(2, 0.62)};
    \addplot[fill=plotColor2Light, draw=white!15!black, forget plot] coordinates {(3, 0.1149)};
    \addplot[fill=plotColor2Light, draw=white!15!black, forget plot] coordinates {(4, 0.02)};
    \addplot[fill=plotColor3Light, draw=white!15!black] coordinates {(5, 0.03)};
    \addlegendentry{AF}
    \addplot[fill=plotColor3Light, draw=white!15!black] coordinates {(6, 0.02)};
    \addplot[fill=plotColor3Light, draw=white!15!black] coordinates {(7, 0.0085)};
\end{axis}

\begin{axis}[
    width=\fwidth,
    height=\fheight,
    at={(0\fwidth,0\fheight)},
    scale only axis,
    legend image post style={mark indices={}},
    legend style={
        /tikz/every even column/.append style={column sep=0.2cm},
        at={(0.5, 1.03)}, 
        anchor=south, 
        draw=white!80!black, 
        font=\scriptsize
        },
    legend columns=3,
    xlabel style={font=\footnotesize},
    xtick align=inside,
    ytick align=inside,
    xlabel={Relay configuration},
    xtick={0,1,2,3,4,5,6,7},
    xticklabels={gNB, 10x20, 20x40, 40x80, 60x120, 4x4, 8x8, 16x16},
    xmajorgrids,
    xmin=-0.5, xmax=7.5,
    xtick style={color=white!15!black},
    ylabel shift = -1 pt,
    ylabel style={font=\footnotesize, align=center},
    ylabel={Average PER},
    ymajorgrids,
    ymin=0, ymax=1,
    ytick style={color=white!15!black},
    ybar,
    bar width=.3,
    bar shift=0pt,
    legend image code/.code={
        \draw [#1] (0cm,-0.1cm) rectangle (0.3cm, 0.25cm); }
]


    \addplot[fill=plotColor1, draw=white!15!black, postaction={pattern={dots}}] coordinates {(0, 1)};
    \addlegendentry{gNB only}

    \addplot[fill=plotColor2, draw=white!15!black, postaction={pattern=north east lines}] 
        coordinates {(1, 0.99)};
    \addlegendentry{IRS}
    \addplot[fill=plotColor2, draw=white!15!black, forget plot, postaction={pattern=north east lines}]  coordinates {(2, 0.8442)};
    \addplot[fill=plotColor2, draw=white!15!black, forget plot, postaction={pattern=north east lines}]  coordinates {(3,0.51)};
    \addplot[fill=plotColor2, draw=white!15!black, forget plot, postaction={pattern=north east lines}]  coordinates {(4, 0.224)};
    \addplot[fill=plotColor3, draw=white!15!black] coordinates {(5, 0.6865)};
    \addlegendentry{AF}
    \addplot[fill=plotColor3, draw=white!15!black] coordinates {(6, 0.3)};
    \addplot[fill=plotColor3, draw=white!15!black] coordinates {(7, 0.0362)};
    
\end{axis}

\end{tikzpicture}